\documentstyle[prd,tighten,aps]{revtex}
\begin{document}
\draft

\twocolumn[\hsize\textwidth\columnwidth\hsize\csname
@twocolumnfalse\endcsname
\renewcommand{\theequation}{\thesection . \arabic{equation} }
\title{\bf On the warp drive space-time}

\author{Pedro F. Gonz\'{a}lez-D\'{\i}az}
\address{Centro de F\'{\i}sica ``Miguel Catal\'{a}n'',
Instituto de Matem\'{a}ticas y F\'{\i}sica Fundamental,\\ Consejo
Superior de Investigaciones Cient\'{\i}ficas, Serrano 121, 28006
Madrid (SPAIN)}
\date{june 23, 1999}

\maketitle

\begin{abstract}
In this paper the problem of the quantum stability of the
two-dimensional warp drive spacetime moving with an apparent
faster than light velocity is considered. We regard as a maximum
extension beyond the event horizon of that spacetime its
embedding in a three-dimensional Minkowskian space with the
topology of the corresponding Misner space. It is obtained that
the interior of the spaceship bubble becomes then a multiply
connected nonchronal region with closed spacelike curves and
that the most natural vacuum allows quantum fluctuations which
do not induce any divergent behaviour of the re-normalized
stress-energy tensor, even on the event (Cauchy) chronology
horizon. In such a case, the horizon encloses closed timelike
curves only at scales close to the Planck length, so that the
warp drive satisfies the Ford's negative energy-time inequality.
Also found is a connection between the superluminal
two-dimensional warp drive space and two-dimensional
gravitational kinks. This connection allows us to generalize the
considered Alcubierre metric to a standard, nonstatic metric
which is only describable on two different coordinate patches.
\end{abstract}


\vskip2pc]

\renewcommand{\theequation}{\arabic{section}.\arabic{equation}}

\section{\bf Introduction}
\setcounter{equation}{0}

General relativity admits many rather unexpected solutions most
of which represent physical situations which have been thought
to be pathological in a variety of respects, as they correspond
to momentum-energy tensors which violate classical conditions
and principles considered as sacrosant by physicists for many
years [1]. Among these solutions you can find Lorentzian
wormholes [2], ringholes [3], Klein bottleholes [4], the
Gott-Grant's double-string [5,6], the Politzer time machine [7],
the multiply connected de Sitter space [8,9], a time machine in
superfluid $^{3}$He [10], etc, all of which allow the existence
of closed timelike curves (CTC's), so as spacetimes where
superluminal, though not into the past travels are made
possible, such as it happens in the Alcubierre warp drive [11].
Most of the solutions that contain CTC's are nothing but
particular modifications or generalizations from Misner space
which can thereby be considered as the prototype of the
nonchronal pathologies in general relativity [12].

Time machines constructed from the above mentioned spacetimes
contain nonchronal regions that are generated by shortcutting
the spacetime and can allow traveling into the past and future
at velocities that may exceed the speed of light. A solution to
Einstein equations that has also fascinated and excited
relativists is the so-called warp drive spaceime discovered by
Alcubierre [11]. What is violated in this case is only the
relativistic principle that a space-going traveller may move
with any velocity up to, but not including or overcoming, the
speed of light. Alcubierre's construct corresponds to a warp
drive of science fiction in that it causes spacetime to contract
in front of a spaceship bubble and expand behind, so providing
the spaceship with a velocity that can be much greater than the
speed of light relative to distant objects, while the spaceship
never locally travels faster than light. Thus, the ultimate
relativistic reason making possible such a warp drive within the
context of general relativity is that, whereas two observer at
the same point cannot have relative velocities faster than
light, when such observers are placed at distant locations their
relative velocity may be arbitrarily large while the observers
are moving in their respective light cones.

The properties of a warp drive and those of spacetimes with
CTC's can actually be reunited in a single spacetime. Everett
has in fact considered [13] the formation of CTC's which arise
in the Alcubierre warp drive whenever it is modified so that it
can contain two sources for the gravitational disturbance which
are allowed to move past one another on parallel, noncolinear
paths. The resulting time machine can then be regarded to
somehow be a three-dimensional analog of the two-dimensional
Gott's time machine [5] in which the CTC's encircle pairs of
infinite, stright, parallel cosmic strings which also move
noncolinearly. In the present paper we consider the physical and
geometric bases which allow us to construct a two-dimensional
single-source Alcubierre warp drive moving at constant faster
than light apparent speed, that at the same time can be viewed
as a time machine for the astronaut travelling in the spaceship.
This can be achieved by converting the interior of the spaceship
bubble in a multiply connected space region satisfying the
identification properties of the three-dimensional Misner space
[14] . We shall also show how in doing so some of the acutest
problems of the Alcubierre warp drive can be solved.

It was Alcubierre himself who raised [11] doubts about whether
his spacetime is a physically reasonable one. Energy density is
in fact negative within the spaceship bubble. This may in
principle be allowed by quantum mechanics, provided the amount
and time duration of the negative energy satisfy Ford like
inequalities [15]. However, Ford and Pfenning have later shown
[16] that even for bubble wall thickness of the order of only a
few hundred Planck lengths, it turns out that the integrated
negative energy density is still physically unreasonable for
macroscopic warp drives. This problem has been recently
alleviated, but not completely solved, by Van Den Broeck [17]
who by only slightly modifying the Alcubierre spacetime,
succeeded in largely reducing the amount of negative energy
density of the warp. A true solution of the problem comes only
about if the warp bubble is microscopically small [16], a
situation which will be obtained in this paper when CTC's are
allowed to exist within the warp bubble.

On the other hand, Hiscock has computed [18] for the
two-dimensional Alcubierre warp drive the stress-energy tensor
of a conformally invariant field and shown that it diverges on
the event horizon appearing when the apparent velocity exceeds
the speed of light, so rendering the warp drive instable. A
similar calculation performed in this paper for the case where
the interior spacetime of the spaceship bubble is multiply
connected yields a vanishing stress-energy tensor even on the
event horizon which becomes then a (Cauchy) chronology horizon,
both for the self consistent Li-Gott vacuum [19] and when one
uses a modified three-dimensional Misner space [20] as the
embedding of the two-dimensional warp drive spacetime. It is
worth noticing that if the astronaut inside the spaceship is
allowed to travel into the past, then most of the Alcubierre's
warp drive problems derived from the fact that an observer at
the center of the warp bubble is causally separated from the
bubble exterior [21] can be circumvented in such a way that the
observer might now contribute the creation of the warp bubble
and control one once it has been created, by taking advantage of
the causality violation induced by the CTC's.

We outline the rest of the paper as follows. In Sec. II we
biefly review the spacetime geometry of the Alcubierre warp
drive, discussing in particular its two-dimensional metric
representation and extension when the apparent velocity exceeds
the speed of light. The visualization of the two-dimensional
Alcubierre spacetime as a three-hyperboloid embedded in $E^{3}$
and its connection with the three-dimensional Misner space is
dealt with in Sec. III, where we also covert the two-dimensional
warp drive spacetime into a multiply connected space by making
its coordinates satisfy the identification properties of the
three-dimensional Misner space. By replacing the Kruskal maximal
extension of the geodesically incomplete warp drive space with
the above embedding we study in Sec. IV the Euclidean
continuation of the multiply connected warp drive, identifying
the periods of the variables for two particular ans\"atze of
interest. We also consider the Hadamard function and the
resulting renormalized stress-energy tensor for each of these
ans\"atze, comparing their results. Finally, we conclude in Sec.
V. Throughout the paper units so that $G=c=\hbar=1$ are used.

\section{\bf The Alcubierre warp drive spacetime}
\setcounter{equation}{0}

The Alcubierre spacetime having the properties associated with
the warp drive can be described by a metric of the form [11]
\begin{equation}
ds^2=-dt^2+\left(dx-vf(r)dt\right)^2 +dy^2+dz^2 ,
\end{equation}
with $v=dx_s(t)/dt$ the apparent velocity of the warp drive
spaceship, $x_s(t)$ the trajectory of the spaceship along
coordinate $x$, the radial coordinate being defined by
\begin{equation}
r=\left\{\left[x-x_s(t)\right]+y^2+z^2\right\}^{\frac{1}{2}},
\end{equation}
and $f(r)$ an arbitrary function subjected to the boundary
conditions that $f=1$ at $r=0$ (the location of the spaceship),
and $f=0$ at infinity.

Most of the physics in this spacetime concentrates on the
two-dimensional space resulting from setting $y=z=0$, defining
the axis about which a cylindrically symmetric space develops;
thus, the two-dimensional Alcubierre space still contains the
entire worldline of the spaceship. If the apparent velocity of
the spaceship is taken to be constant, $v=v_0$, then the metric
of the two-dimensional Alcubierre space becomes [18]:
\begin{equation}
ds^2=-\left(1-v_0^2 f(r)^2\right)dt^2-2v_0 f(r)dtdx+dx^2,
\end{equation}
with $r$ now given by $r=\sqrt{(x-v_0 t)^2}$, which in the past
of the spaceship ($x>v_0 t$) can simply be written as $r=x-v_0
t$. Metric (2.3) can still be represented in a more familiar
form when one chooses as coordinates $(t,r)$, instead of
$(t,x)$. This can obviously be achieved by the replacement
$dx=dr+v_0 dt$, with which metric (2.3) transforms into:
\begin{equation}
ds^2=-A(r)\left[dt-\frac{v_0\left(1-f(r)\right)}{A(r)}dr\right]^2
+\frac{dr^2}{A(r)},
\end{equation}
where the Hiscock function [18]
\begin{equation}
A(r)=1-v_0^2\left(1-f(r)\right)^2
\end{equation}
has been introduced. Metric (2.4) can finally be brought into a
comoving form, in terms of the proper time
$d\tau=dt-v_0\left(1-f(r)\right)dr/A(r)$,
\begin{equation}
ds^2=-A(r)d\tau^2+\frac{dr^2}{A(r)} .
\end{equation}
As pointed out by Hiscock [18], this form of the metric is
manifestly static. The case of most interest corresponds to
apparent velocities $v_0>1$ (superluminal velocity) where the
metrics (2.4) and (2.6) turn out to be singular with a
coordinate singularity (apparent event horizon) at a given value
of $r=r_0$ such that $A(r_0)=0$, i.e. $f(r_0)=1-1/v_0$.

We note that metric (2.4) can be regarded to be the kinked
metric (describing a gravitational topological defect [22]) that
corresponds to the static metric (2.6). To see this, let us
first redefine the coordinates $t$ and $r$ as
$dt'=A^{\frac{1}{4}}dt$ and $dr'=A^{-\frac{1}{4}}dr$, so that
metric (2.4) can be re-written along the real intervals $0\leq
r'\leq\infty$ (i.e. $0\leq r\leq r_0$) and $0\leq t'\leq\infty$
as
\begin{equation}
ds^2=-\sqrt{A(r)}\left((dt')^2-(dr')^2\right) +2v_0(1-f(r))dt'
dr'.
\end{equation}
Now, metric (2.7) can be viewed as the metric describing at
least a given part of a two-dimensional one-kink (gravitational
topological charge $\pm 1$) if we take $\sin(2\alpha)=\pm
v_0(1-f(r))$, i.e. $\cos(2\alpha)=\sqrt{A}$, so that this metric
can be transformed into
\begin{equation}
ds^2=-\cos(2\alpha)\left((dt')^2-(dr')^2\right)\pm
2\sin(2\alpha)dt'dr' ,
\end{equation}
where $\alpha$ is the tilt angle of the light cones tipping over
the hypersurfaces [22,23], and the choice of sign in the second
term depends on whether a positive (upper sign) or negative
(lower sign) gravitational topological charge is considered.

The existence of the complete one-kink is allowed whenever one
lets $\alpha$ to monotonously increase from 0 to $\pi$, starting
with $\alpha(0)=0$, with the support of the kink being the
region inside the event horizon. Then metric (2.8) converts into
metric (2.6) if we introduce the substitution
$\sin\alpha=\sqrt{\left(1\pm\sqrt{A(r)}\right)/2}$ and the
change of time variable $\tau=t'+g(r)$, with
$dg(r)/dr'=\tan(2\alpha)$. The region inside the warp drive
bubble supporting the kink is the only region which can actually
be described by metric (2.8) because $\sin\alpha$ cannot exceed
unity and hence $0\leq r\leq r_0$. In order to have a comple
description of the one-kink and therefore of the warp drive one
need a second coordinate patch where the other half of the
$\alpha$ interval, $\pi/2\leq\alpha\leq\pi$, can be described.
This can be achieved by introducing a new time coordinate [23]
$\bar{t}'=t'+h(r)$, with
$dh(r)/dr=\left[v_0(1-f(r))-k\right]/A^{\frac{3}{4}}$, in which
$k=\pm 1$, the upper sign for the first patch and the lower one
for the second patch. This choice is adopted for the following
reason. The zeros of the denominator of
$dh/dr'=\left(\sin(2\alpha)\mp 1\right)/\sqrt{A}$ correspond to
the two event horizons where $r=r_0$ and $f(r_0)=1-1/r_0$, one
per patch. For the first patch, the horizon occurs at
$\alpha=\pi/4$ and therefore the upper sign is selected so that
both $dh/dr$ and $h$ remain well defined and the kink is
preserved on this horizon. For the second patch the horizon
occurs at $\alpha=3\pi/4$ and therefore the lower sign is
selected for it. The two-dimensional metric for a (in this
sense) complete warp drive will be then [23]
\begin{equation}
ds^2=-A(r)d\bar{t}^2\mp 2kd\bar{t}dr ,
\end{equation}
or in terms of the $(\bar{t},x)$ coordinates in the
four-dimensional manifold,
\begin{equation}
ds^2=-\left[A(r)\mp 2kv_0\right]d\bar{t}^2\mp 2kd\bar{t}dx
+dy^2+dz^2 ,
\end{equation}
with $r$ as defined by Eq. (2.2).

Having shown the existence of a connection between warp drive
spacetime and topological gravitational kinks and hence extended
the warp drive metric to that described by Eq. (2.10), we now
return to analyse metric (2.6) by noting that the geodesic
incompleteness of this metric at $r=r_0$ can, as usual, be
avoided by maximally extending it according to the Kruskal
technique. For this to be achieved we need to define first the
quantity $r^{*}=\int\frac{dr}{A(r)}$. Using for $f(r)$ the
function suggested by Alcubierre
\begin{equation}
f(r)=
\frac{\tanh\left[\sigma(r+R)\right]-\tanh\left[\sigma(r-R)\right]}
{2\tanh(\sigma R)} ,
\end{equation}
where $R$ and $\sigma$ are positive arbitrary constants, we
obtain
\[r^{*}=\frac{(1+2v_0)r}{1-v_0^2}\]
\[+\frac{(3+2v_0-v_0^2)v_0^2}{\sigma(1-v_0^2)^{2}\sqrt{3v_0^2+2v_0-1}}\]
\[\times\ln\left[\frac{2v_{0}(1+v_0)\tanh(\sigma r)-\sqrt{3v_0^2+2v_0-1}}{2v_0(1+v_0)\tanh(\sigma r)+\sqrt{3v_0^2+2v_0-1}}\right] \]
\[-\frac{v_{0}(1+3v_0)}{4\sigma(1+v_0)\sqrt{2v_{0}(1+v_0)}}\]
\[\times\ln\left[\frac{1+v_{0}-\sqrt{2v_{0}(1+v_0)}\tanh(\sigma r)}{1+v_{0}+\sqrt{2v_{0}(1+v_0)}\tanh(\sigma r)}\right] \]
\begin{equation}
+\frac{v_0}{2\sigma(1+v_0)}\tanh(\sigma r) ,
\end{equation}
where we have specialized to the allowed particular case for
which $\sinh^{2}(\sigma R)=v_0$. Introducing then the usual
coordinates $V=t+r^{*}$ and $W=t-r^{*}$, so that
\begin{equation}
ds^2= -\left[1-\frac{v_{0}^2\sinh^4(\sigma
r)}{\left(\cosh^2(\sigma r)+v_0\right)^2}\right]dVdW ,
\end{equation}
and hence the new coordinates
\[\tanh V'=\exp\left(\frac{\sigma
V}{4\sinh^{-1}\sqrt{\frac{v_{0}+1}{v_{0}-1}}}\right) \]

\[\tanh W'=-\exp\left(\frac{-\sigma
W}{4\sinh^{-1}\sqrt{\frac{v_{0}+1}{v_{0}-1}}}\right), \] we
finally obtain the maximally extended, geodesically complete
metric
\[ds^2=\]
\begin{equation}
\frac{64}{\sigma^2}\left[1-\frac{v_0^2\sinh^4(\sigma
r)}{\left(\cosh^2(\sigma
r)+v_0\right)^2}\right]\frac{\left(\sinh^{-1}\sqrt{\frac{v_{0}+1}{v_{0}-1}}\right)^{2}dV'
dW'}{\sin(2V')\sin(2W')} ,
\end{equation}
where $r$ is implicitly defined by
\begin{equation}
\tan V'\tan W'=-\exp\left(\frac{\sigma
r^{*}}{2\sinh^{-1}\sqrt{\frac{v_0+1}{v_0-1}}}\right) .
\end{equation}
The maximal extension of metric (2.6) is thus obtained by taking
expression (2.14) as the metric of the largest manifold which
metrics given only either in terms of $V$ or in terms of $W$ can
be isometrically embedded. There will be then a maximal manifold
on which metric (2.14) is $C^2$ [24].

\section{\bf Warp drive with internal CTC's}
\setcounter{equation}{0}

In this section we investigate a property of the two-dimensional
Alcubierre spacetime which will allow us to avoid the
complicatedness of Kruskal extension in order to study its
Euclidean continuation and hence its stability against quantum
vacuum fluctuations. Thus, taking advantage from the similarity
between metric (2.6) and the de Sitter metric in two dimensions,
we can visualize the dimensionally reduced Alcubierre spacetime
as a three-hyperboloid defined by
\begin{equation}
-v^2+w^2+x^2=v_0^{-2} ,
\end{equation}
where $v_0 >1$. This hyperboloid is embedded in $E^3$ and the
most general expression for the two-dimensional metric of
Alcubierre space for $v_0 >1$ is then that which is induced in
this embedding, i.e.:
\begin{equation}
ds^2=-dv^2+dw^2+dx^2 ,
\end{equation}
which has topology $R\times S^2$ and invariance group $SO(2,1)$.

Metric (3.2) can in fact be conveniently exhibited in static
coordinates $\tau\in(-\infty,\infty)$ and $r\in(0,r_0)$, defined
by
\[
v=v_0^{-1}\sqrt{A(r)}\sinh(v_0 \tau) \]
\begin{equation}
w=v_0^{-1}\sqrt{A(r)}\cosh(v_0 \tau)
\end{equation}
\[x=F(r) ,\]
where
\begin{equation}
\left[F(r)'\right]^2=-\left[\frac{\left(\frac{d\rho}{dr}\right)^2-
4v_0^2}{4v_0^2 (1+\rho)}\right] ,
\end{equation}
the prime denoting derivative with respect to radial coordinate
$r$, with
\begin{equation}
\rho\equiv\rho(r)=-v_0 \left(1-f(r)\right)^2
\end{equation}
and
\begin{equation}
A=1+\rho(r) .
\end{equation}
Using coordinates (3.3) in metric (3.2) we re-derive then metric
(2.6). Thus, the two-dimensional Alcubierre space can be
embedded in Minkowski space in three dimensions. Since Minkowski
metric (3.2) plus the identifications
\[(v,w,x)\leftrightarrow\]
\begin{equation}
\left(v\cosh(nb)+w\sinh(nb),v\sinh(nb)+w\cosh(nb),x\right),
\end{equation}
where $b$ is a dimensionless arbitrary boosting quantity and $n$
is any integer number, make the universal covering [24] of the
Misner space in three dimensions, one can covert the
two-dimensional Alcubierre space into a multiply connected space
if we add the corresponding identifications
\begin{equation}
(\tau,r)\leftrightarrow (\tau+\frac{nb}{v_0},r)
\end{equation}
in the original warp drive coordinates. The boost transformation
in the $(v,w)$ plane implied by identifications (3.7) will
induce therefore the boost transformation (3.8) in the
two-dimensional Alcubierre space. Hence, since the boost group
in Alcubierre space must be a subgroup of the invariance group
of the two-dimensional Alcubierre embedding, the static metric
(2.6) can also be invariant under symmetry (3.7). Thus, for
coordinates defined by Eqs. (3.3) leading to the static metric
with an apparent horizon as metric (2.6), the symmetry (3.7) can
be satisfied in the region covered by such a metric, i.e. the
region $w>|v|$, where there are CTC's, with the boundaries at
$w=\pm v$, and $x^2=v_0^{-2}$ being the Cauchy horizons that
limit the onset of the nonchronal region from the Alcubierre
causal exterior. Such boundaries are situated at $r_0$, defined
by $f(r_0)=1-v_0^{-1}$, and become then appropriate chronology
horizons [12] for the so-obtained multiply connected
two-dimensional Alcubierre space with $v_0 >1$.

We have in this way succeeded in coverting a two-dimensional
warp drive with constant, faster than light apparent velocity in
a multiply connected warp drive with CTC's only inside the
spaceship, and its event horizon at $r_0$ in a chronology
horizon. This is a totally different way of transforming warp
drives into time machines of the mechanism envisaged by Everett
[13] for generating causal loops using two sources of
gravitational disturbance which move past one another. In our
case, even though the astronaut at the center of the warp bubble
is still causally separated from the external space, he (or she)
can always travel into the past to help creating the warp drive
on demand or set up the initial conditions for the control of
one once it has been created.

\section{\bf Quantum stability of multiply connected warp drive}
\setcounter{equation}{0}

In this section we shall show that the two-dimensional multiply
connected warp drive spacetime is perfectly stable to the vacuum
quantum fluctuations if either a self consistent Rindle vacuum
is introduced, or for microscopic warp bubbles. We have already
shown that the two-dimensional warp drive spacetime can be
embedded in the Minkowskian covering of the three-dimensional
Misner space when the symmetries of this space implied by
identifications (3.7) (that lead to identifications (3.8) in
Alcubierre coordinates) are imposed to hold also in the
two-dimensional Alcubierre spacetime with $v_0 >1$. Since the
embedding can be taken to play an analogous role to that of a
maximal Kruskal extension, it follows that showing stability
against vacuum quantum fluctuations in three-dimensional Misner
space would imply that the two-dimensional multiply connected
warp drive space with $v_0 >1$ is also stable to the the same
fluctuations. This conclusion is in sharp contrast with some
recent result obtained by Hiscock who has shown [18] that the
stress-energy tensor for a conformally invariant scalar field
propagating in simply connected two-dimensional Alcubierre space
diverges on the event horizon if the apparent velocity of the
spaceship exceeds the speed of light. He obtained an observed
energy density near the horizon proportional to
$\left[f-(1-v_0^{-1})\right]^{-2}$, which in fact diverges as
$r\rightarrow r_0$.

Metric (3.2) can be transformed into the metric of a
three-dimensional Misner space explicitly by using the coodinate
re-definitions
\[v=\theta\cosh x^1 \]
\begin{equation}
w=\theta\sinh x^1
\end{equation}
\[ x=x^2 .  \]
Then,
\begin{equation}
ds^2=-d\theta^2+\theta^{2}\left(dx^1\right)^2+\left(dx^2\right)^2
,
\end{equation}
which is the three-dimensional Misner metric for coordinates
$0<\theta<\infty$, $0\leq x^1\leq 2\pi$ and $0\leq x^2
\leq\infty$. This metric is singular at $\theta=0$ and, such as it
happens in its four-dimensional extension, it has CTC's in the
region $\theta<0$. Note that one can also obtain the line
element (4.2) directly from the two-dimensional Alcubierre
metric with $v_0 >1$ by introducing the coordinate
transformations
\begin{equation}
\theta=\frac{i\sqrt{A(r)}}{v_0}
\end{equation}
\begin{equation}
x^1=-i\arcsin\left[\cosh(v_0\tau)\right]
\end{equation}
and
\begin{equation}
x^2=F(r),
\end{equation}
with $F(r)$ as defined by Eq. (3.4). Let us now consider the
Euclidean continuation from which metric (4.2) becomes positive
definite. This is accomplished if we rotate both coordinates
$\theta$ and $x^1$ simultaneously, so that
\begin{equation}
\theta=i\eta ,\;\;\; x^1=i\chi ,
\end{equation}
where $\eta$ and $\chi$ can be expressed in terms of the
Alcubierre coordinates $r$ and $\tau$ by means of
transformations (4.3) and (4.4). The covering space of the
resulting metric preserves however the Lorentzian signature.
This might be an artifact coming from the singular character of
metric (4.2), so it appears most appropriate to extend first
this metric beyond $\theta=0$ in order to get the Euclidean
sector of the three-dimensional Misner metric, and hence
investigate the stability of the superluminal, multiply
connected two-dimensional warp drive spacetime against vacuum
quantum fluctuations. Extension beyond $\theta=0$ of metric
(4.2) is conventionally made by using new coordinates [24]
$T=\theta^2$, $V=\ln\theta+x^1$ and $x^2=x^2$, so that
\begin{equation}
ds^2=-dTdV+TdV^2+\left(dx^2\right)^2 ,
\end{equation}
or by re-defining $V=Y+Z$ and $T-\int TdV=Y-Z$,
\begin{equation}
ds^2=-dY^2+dZ^2+\left(dx^2\right)^2
\end{equation}
and
\begin{equation}
Y^2-Z^2=V\left(T-\int TdV\right) ,\;\;
\frac{Y-Z}{Y+Z}=\frac{t-\int TdV}{V} .
\end{equation}
Metric (4.8) becomes positive definite when continuing the new
coordinate $Y$ so that $Y=i\zeta$. By using this continuation
together with rotations (4.6) in expressions (4.9), we can
deduce that the section on which $\zeta$ and $Z$ are both real
corresponds to the region defined by $\theta\geq
e^{\frac{1}{2}}$, $x^1\geq 1/2$, and that there exist two
possible ans\"atze according to which the continuation can be
performed. One can first set
\[\exp(i\zeta)= \]
\begin{equation}
i\exp\left(\frac{\zeta^2
-Z^2}{Z}\right)\eta\exp\left[-2\chi\left(\chi+
\frac{\pi}{2}\right)\right]\exp(i\chi) ,
\end{equation}
where only coordinate $\chi$ turns out to be periodic on the
Euclidean sector, with a period $\bar{b}=2\pi$. Rotating back to
the Lorentzian region, we then have $b=2\pi$. Ansatz (4.10)
should be associated with the self consistent Rindler vacuum
considered by Li and Gott for four-dimensional Misner space
[19], instead of the Minkowski vacuum with multiple images
originally used by Hiscock and Konkowski [25]. Introducing then
Rindler coordinates defined by $v=\xi\sinh\omega$ and
$w=\xi\cosh\omega$ in the three-dimensional covering metric
(3.2), so that it becomes
\[ ds^2=-\xi^{2}d\omega^2 +d\xi^2 +dx^2, \]
we can compute the Hadamard function for a conformally invariant
scalar field in Rindler vacuum to be
\begin{equation}
G^{(1)}(X,X')=
\frac{1}{2\pi^2}\frac{\gamma}{\xi\xi'\sinh\gamma\left[-(\omega-
\omega')^2 +\gamma^2\right]} ,
\end{equation}
with $X=(\omega,\xi,x)$, $X'=(\omega',\xi',x')$, and
\begin{equation}
\cosh\gamma=\frac{\xi^2+\xi'^{2}+(x-x')^2}{2\xi\xi'} .
\end{equation}
Using the method of images [25] and the usual definitions of the
regularized Hadamard function and hence the renormalized
stress-energy tensor [18], we finally get for the latter
quantity an expression which is proportional to
\[ \frac{1}{\xi^3}\left[\left(\frac{2\pi}{b}\right)^3 -1\right]
,\] which, if we take $b=2\pi$ according to the above ansatz,
obviously vanishes everywhere, even on the (Cauchy) chronology
horizon at $\xi=0$.

Although, in spite of the Ford-Pfenning's requirement [16], this
is allowing the existence of stable superluminal, multiply
connected warp drives of any size, this choice for the vacuum
has two further problems. First of all, previous work by Kay,
Radzikowski and Wald [26] and by Cassidy [27] casts compelling
doubts on the meaning of the Cauchy horizon, and hence on the
validity of the conclusion that the stress-energy tensor is zero
also on such a horizon. On the other hand, having an Euclidean
section on which time is not periodic (which implies
nonexistence of any background thermal radiation) while the
spacetime has an event horizon appears to be rather
contradictory. It is for these reasons that we tend to favour
the second possible ansatz implementing the Euclidean
continuation of the three-dimensional Misner covering, which
appears to be less problematic. It reads:
\[\exp\left(\frac{i\xi}{Z}\right)=\exp\left(\frac{\zeta^2
-Z^2}{2Z^2}\right)\eta^{\eta^{-2}}\times \]
\begin{equation}
\exp\left[-\frac{2\chi\left(\chi
+\frac{\pi}{2}\right)}{\eta^2}\right]\exp\left[i\left(\frac{\chi
+\frac{\pi}{2}}{\eta^2}\right)\right] ,
\end{equation}
where both the Euclidean time $\eta$ and the Euclidean
coordinate are now periodic, with respective dimensionless
periods $\Pi_{\eta}=1/2$ and $\Pi_{\chi}=2\pi\eta^2$. The
physical time period in two-dimensional Alcubierre space at
faster than light apparent velocity, $\Pi_{Al}$, can be related
to period $\Pi_{\eta}$ by means of the expression
\[
v_{0}\Pi_{Al}=
\left.4\pi\Pi_{\eta}\sqrt{g_{Al}^{rr}}\frac{dr}{d\eta}\right|_{r=r_0}
.\] Using then the Euclidean continuation of Eq. (4.3), one can
obtain $\Pi_{Al}=4\pi/A(r_{0})'$, which corresponds to a
background temperature, $T_{Al}=A(r_{0})'/4\pi$, that is the
same as that which is associated with the event horizon of the
spaceship and was first derived by Hiscock [18]. Rotating back
to the Lorentzian section we see that time $\theta$ becomes
again no longer periodic, but $x^1$ keeps still a periodic
character with period $b=2\pi\theta^2$. If we adhere to ansatz
(4.13), this would mean that the Misner space itself should be
modified in such a way that its spatial volume would vanish as
time $\theta$ approaches zero [20]. When calculating then the
regularized Hadamard function, one should use a method of images
which ought also to be modified accordingly with the fact that
the period of the closed spatial direction is time dependent. If
we use a most general automorphic field and impose constancy for
the frequency of the general solution of the wave equation [20],
one is led [20] to a time quantization that unavoidably implies
an also strictly zero value for the renormalized stress-energy
tensor, everywhere in the whole two-dimensional, superluminal
Alcubierre space. The price to be paid [20] for this quantum
stability is that the CTC's developing inside the warp bubble
and actually the warp bubble itself should never exceed a
submicroscopic size near the Planck scale, so avoiding not just
the unwanted Hiscock instability, but also any violation of
negative energy-time Ford like [15,16] inequality and hence any
unphysical nature of the warp drive. A possible remaining
question is whether the CTC's within the bubble might produce
new divergences, at least if Hawking's chronology protection
conjecture [28] is correct. However, the semiclassical
instabilities leading to chronology protection are actually of
the kind which are precisely prevented in our above model.

\section{\bf Summary and conclusions}
\setcounter{equation}{0}

Among the achronal pathologies which are present in general
relativity and that can be associated with the symmetries of the
Misner space, we consider in this paper the two-dimensional warp
drive with an apparent velocity which is faster than light, and
whose spaceship interior is multiply connected and therefore
nonchronal. After reviewing the geometrical properties of the
Alcubierre-Hiscock two-dimensional model, it has been proved
that there exists a close connection between warp drives and
gravitational kinky topological defects, at least in two
dimensions. Generalizing to the standard Finkenstein kinked
metric to allow a complete description of the one-kinks, we also
generalize the Alcubierre metric in such a way that it can no
longer be described in just one coordinate patch.

The geodesic incompleteness of the Alcubierre-Hiscock space at
the event horizon for superluminal warp bubbles has been
eliminated by first extending the metric beyond the horizon
according to the Kruskal procedure, and then by an embedding in
a three-dimensional Minkowski space. It has been also shown
that, if the latter space is provided with the topological
identifications that correspond to the universal covering of the
three-dimensional Misner space, then one can convert the
interior of the warp spaceship into a multiply connected space
with closed timelike curves which is able to behave like a time
machine.

The problem of the quantum stability of the two-dimensional,
multiply connected warp drive spacetime has been finally
considered. Using the three-dimensional Misner embedding as the
maximal extension of the two-dimensional warp drive space, it
has also been shown that the divergence encountered by Hiscock
on the event horizon for the simply connected case is smoothed
out, while the apparent horizon becomes a regularized (Cauchy)
chronology horizon. We argued as well that the unphysical
violation of the negative energy-time inequalities can at the
same time be circumvented because the most consistent quantum
treatment for dealing with vacuum fluctuations in the multiply
connected case leads also to the result that the size of the
spaceship bubble and its closed timelike curves must necessarily
be placed at scales close to the Planck length.

\acknowledgements
The author thanks C.L. Sig\"uenza for useful discussions. This
work was supported by DGICYT under Research Project No.
PB97-1218.

\end{document}